\documentclass[10pt]{article}
\begin{document}

\title{Cosmology and Local Physics}
\author{George F R Ellis}
\maketitle
\begin{abstract}
This article is dedicated to the memory of Dennis Sciama. It revisits
a series of issues to which he devoted much time and effort, regarding the relationship between local physics and the large scale structure of the universe - in particular, Olber's paradox, Mach's principle, and the various arrows of time. Thus the focus is various ways in which local physics is influenced by the universe itself.
\end{abstract}

\section{Introduction}

The major thrust of present day scientific cosmology is that of examining
the effect of local physical laws on the large-scale structure of the
cosmos. This started with Einstein's application of the local law of
gravitation, expressed via the Einstein field equations, to determine
space-time structure in the large \cite{einstein,hawell}. This then led to
the Friedmann-Lema\^{\i}tre-Eddington demonstration of how these equations
imply an evolution of the universe \cite{history}, followed by the
prediction, on the basis of classic gravitational theory, of a beginning to
the universe at a space-time singularity \cite{sing,ehlers,hawell,tce}. This
thrust attained new power through the understanding of how nuclear physics
processes in the hot early expansion phase of the universe would lead to
synthesis of light elements out of primordial constituents, providing an
explanation both for the origin of the light elements and the existence of
the 3K blackbody Cosmic Blackbody Radiation (`CBR') \cite{nucleo} - a theme
that Dennis Sciama found most exciting, and contributed to in his inimitable
way  \cite{sciabk,scia-varenna}. And it has reached its climax in the
understanding, following Alan Guth's innovative realisation of the power of
the idea of an inflationary era of expansion in the early universe caused by
energy-condition violating quantum fields \cite{guth}, of how particle
physics processes in the very early universe could be of major importance in
determining the large-scale structure of the universe today \cite
{inflate1,inflate2}.

However there has also from the earliest times been a counter theme: the
study of the way that global properties of the universe can influence its
local properties. This was particularly embodied in Mach's principle, and
was always dear to Dennis' heart, inspired particularly by his discussions
with Fred Hoyle, Tommy Gold, and Herman Bondi when they were together in
Cambridge in the 1950's. He devoted much thought to this topic, resulting in
his famous vector gravitational model \cite{scia1} and various popular books 
\cite{scia3,scia2}. In his writings on these topics, he consistently
emphasized the \emph{interconnectedness of the universe }\cite{scia3}:\ each
part interacting with each other part, and with very distant parts being as
important as local regions in these interactions, a prime example being
Olber's paradox \cite{bondi}. Because of this interconnection, in principle
one can obtain some understanding of the whole from any part; indeed if one
was clever enough and understood enough physics, one could in principle
completely deduce the nature of the whole from a sufficient study of its
parts. An example of this line of argument is the suggestion that one could
in principle deduce the expansion of the universe, and even the approximate
value of the Hubble constant, from the existence of bus tickets.

A major further theme is the \emph{uniqueness of the universe, }\ which is
what gives cosmology its specific unique nature as a science \cite
{scia3,ellunique}. Because of this uniqueness, one runs into major problems
in distinguishing boundary conditions from physical laws. What appears to be
an inviolable physical law may just be a consequence of the particular
boundary conditions that happen to hold in this particular universe. This
might even be true for some of the apparently most fundamental of laws at
the macroscopic scale, such as the second law of thermodynamics and its
associated arrows of time; but we cannot test this supposition, for we
cannot re-run the universe, nor can we investigate any other universe.
Conversely, one can propose that if the universe were different, the laws of
physics would be different. This leads to the idea that as the universe is
changing, maybe the laws of nature are changing with it, or at least this
might be true for the values of some of the `fundamental constants' of
physics, for example the gravitational constant \cite{dirac}. From this line
of argument follows the need to consider seriously alternative physics as
well as alternative boundary conditions when we consider the relation
between local physics and cosmology.

Many years ago, Dennis and I jointly wrote a survey article on these
global-to-local relations \cite{ellsci}.The purpose of the present paper is
to revisit these topics, and comment on what progress has been made in the
intervening years. I\ deal in turn with Olber's paradox; Mach's principle;
the arrow of time; the framework for relating local to global physics; and
the relation between initial conditions and local physics. This paper is
dedicated to the memory of Dennis, who was an inspiring teacher,
enthusiastic colleague, and good friend \footnote{%
See Roger Penrose's article, pages 314-315 in \cite{sci4}, for a nice account
both of Dennis' infectious passion for physics, and his belief in the
importance of non-local effects.}.

\section{Olber's Calculation}

The basic point of Olber's paradox (`Why is the sky dark at night?'),
actually developed by Halley, is the need for integration over all very
distant sources in determining the intensity of integrated radiation from
astronomical sources. These distant sources cannot be neglected in flat
space-time, because their number goes up as $r^2$ and this compensates for
the decrease in flux from each source, which goes down like $1/r^2$, where $r
$ is the standard radial distance in Minkowski space. Doing the calculation
in Robertson-Walker universes with $r$ chosen as the area distance gives the
same result, and the reciprocity theorem \cite{ell71} shows this will in
fact be true in any curved space-time, for it is equivalent to the result
that total intensity $I$ of radiation received from a source (that is, flux
per unit solid angle) is independent of the source's area distance; it will
vary as

\[
I=\frac 1{(1+z)^4}I_G 
\]
where $I_G$ its bolometric surface brightness, and $z$ its redshift, while
the pointwise specific intensity $I_\nu $ of radiation received (the
intensity per unit frequency range) in direction $\theta ,\phi $ is

\begin{equation}
I_\nu (\theta ,\phi )=\frac{\mathcal{I}(\nu (1+z))}{(1+z)^3}I_G(\theta ,\phi
)  \label{specint}
\end{equation}
where $\mathcal{I}(\nu )$ is the source spectrum and $I_G(\theta ,\phi )$
its surface brightness in the direction of observation. Hence in an
expanding universe, very distant sources at high $z$ appear much fainter
than nearby ones, and so the sky can appear dark even though every line of
sight eventually intersects a source of some kind.

However, as emphasized by Ted Harrison \cite{Harrison}, that is not the
whole story: in the end, the dark night sky is the result of the fact that
there is not enough radiation in the universe to generate a bright night
sky. Most of the lines of sight from the earth effectively end up on the
surface of last scattering of radiation after the hot big bang, because
although there are numerous intervening galaxies, they are mainly
transparent (this is clear from both lensing studies and QSO Lyman forest
observations).

\subsection{Discrete Sources and Background radiation}

Thus the present day astronomical version of the Olber's studies consists of
two parts. The first is the study of the integrated radiation from all
individually unresolvable sources, resulting in a predicted spectrum of
background radiation at all wavelengths coming from these sources, in
particular, optical, radio, and X-ray backgrounds. The detailed theory of
this background then resides in detailed astrophysical speculations about
these sources and their evolution \cite{backg-radn}; observations are
incomplete, because of galactic and inter-galactic absorption.

The second part is a dominant theme of present day theoretical and
observational cosmology: namely prediction and observation of the CBR
spectrum and anisotropies, resulting from us seeing the point by point
brightness of the surface of last scattering of this radiation \cite
{inflate2}. Observations provide an angular power spectrum, and theory
relates that to the formation of structure in the expanding universe;
comparison of theoretical predictions with the observations then enable us
to get tight limits on cosmological parameters \cite{cbr}. Thus this can
legitimately be regarded as the culmination of Olber's investigations into
the integrated radiation emitted by all sources in the universe. The source
here is the uniformly distributed matter at the surface of last scattering,
which at later times will either be incorporated into galaxies or will
reside in intergalactic space. It is the furthest matter we can see by any
kind of electromagnetic radiation, because the universe is opaque at earlier
times, and hence this matter comprises the visual horizon \cite{vishor}.
This is the proper endpoint of the integral in the Olber's calculation, for
we cannot receive radiation from more distant matter. The temperature of
this surface is about $T_e=3000K$, and it emits black body radiation at that
temperature because the matter and radiation are very close to equilibrium
at that epoch; by Eqn.(\ref{specint}) that radiation is received as
blackbody radiation at a temperature $T_o=T_e(1+z)^{-1}$ where $z$ is the
redshift of emission. Because of the redshift factor of about $1000$ for
that matter, we receive black body radiation at about $3000K/1000=3K$ from
that surface. There are small variations to this temperature over the sky
due to gravitational fields and velocity effects, and study of these
anisotropies is a major part of present day cosmology \cite{inflate2,cbr}.

Thus in the real universe, the Olber's integral does not extend to infinity,
and $3K$ is the temperature of the dominant radiation today, and hence is
the temperature we measure for most of the night sky (and for the day sky as
well, away from the sun). That temperature is not due to discrete sources,
but rather is due to the primeval radiation from the hot big bang.

The further new point is the realisation that this calculation is of
considerable importance, apart from its use in helping us determine
cosmological parameters. It implies that the earth is in contact with a
thermal reservoir at a temperature of $3K$, which is the sink into which we
dispose of our excess entropy, generated by all the processes of life. That
sink is essential to the thermodynamic functioning of the biosphere, and
hence to the existence of life on earth, for the biosphere functions by
receiving the black body radiation from the sun, using it to run biological
and atmospheric processes, and then radiating the waste heat away to the sky 
\cite{penrose}. If the temperature of the sky were much higher, and
certainly if it were above $300K$, life like ours on earth would not be
possible. Thus from this viewpoint, the reason we observe the night sky to
be dark is that if that were not true, we would not be here to see it! This
is of course part of the `anthropic universe' argument, briefly touched on
below.

\subsection{The Gravitational Version}

The same issue of course arises in Newtonian gravitational theory, for that
too is an inverse square law. The present-day form of this problem arises in
studies of the large-scale motions of matter induced by the gravitational
field of large-scale inhomogeneities in the universe, such as the `Great
Attractor' (see for example the POTENT series of studies \cite{potent}). In
any particular case, the question here is, has the integral for the
gravitational potential causing the acceleration and hence inducing these
velocities, converged or not? Do we need to consider further-out matter in
our calculation of these integrals?

There are however some major differences from the Olber's calculation.
Firstly, the integral here is carried out in spacelike surfaces rather than
on the past light cone, for it corresponds to the Newtonian limit of the
relativistic equations, where the constraint equations are equivalent to
relations that must be satisfied at constant time (i.e. at each instant) in
the expanding universe. How this is compatible with the relativistic nature
of the field equations was discussed in \cite{ellsci}. In essence:\ these
initial conditions have to satisfy constraint equations, such as the
Hamiltonian constraints, which are usually expressed as `instantaneous'
conditions. This requirement has to be built into the initial conditions for
the universe, which raises significant conceptual problems in some cases.
Once they are satisfied, they will remain satisfied because of the
consistency of the time evolution equations with the constraint equations
(see Maartens \cite{maartens} and Friedrich and Nagy \cite{Friedrich} for
explicit demonstration of this consistency in particular cases).

Secondly, in contrast to the Olber's case just discussed, there is no
natural cut-off to the integral in this case - one must in principle extend
it to spatial infinity, if the universe is indeed spatially homogeneous -
and then it diverges! (this is indeed the reason that Newtonian cosmology
was not arrived at until some decades after relativistic cosmological models
were available). But this is compensated for by the third point: although
the gravitational field in principle diverges, its local effect is of a
vector nature (causing a vectorial acceleration of matter), and hence the
effects from opposite directions cancel.

Thus in effect one renormalizes by omitting the divergent integral due to a
uniform background, and only calculates the remnant effect due to
inhomogeneities superimposed on this uniform background (this is implied in
the rescaling of variables relative to the background expansion that is used
in the standard calculations of peculiar velocities, see for example Peebles 
\cite{peebles}). Thus in this context, `convergence' means continuing the
integral until the effects of all more distant matter are isotropic and
cancel each other out when one does the required vector addition of
resulting gravitational forces. It is assumed here that the further out
matter is indeed isotropically distributed about us on the largest scales;
and this supposition receives some support from the high degree of CBR\
isotropy we measure. An issue that could perhaps still be reflected on is to
what degree the success of the resulting calculations of large scale
velocities can be taken as providing evidence for isotropy of the universe
outside our visual horizon, and perhaps even outside our particle horizon.
This question, alluded to in \cite{ellsci}, remains open. The problem is
that locally we can only measure the total integrated effect on the motion
of matter, and it does not seem that one can separate out the contributions
to the effect from shells of matter around us at different distances.

\section{Mach's Principle}

This discussion leads naturally on to Mach's famous conjecture, largely
motivated by his position in the philosophical debates about relative versus
absolute motion, that the origin of inertia is interactions with very
distant matter \cite{scia2,scia3,bondi}. This issue remains as open today as
ever.

On the one hand, there are a variety of `Anti-Machian' solutions of
Einstein's Field Equations (EFE) - non-singular solutions of the vacuum
equations \cite{anti-mach} - suggesting to many that these equations do not
by themselves incorporate Mach's principle, despite Einstein's hope that
this would be so \cite{Wheeler}. This kind of view is supported by
cosmological analyses showing that distant galaxies are at rest in the local
inertial rest-frame iff the cosmological vorticity is very close to zero 
\cite{schucking,ehlers} - and there clearly are both Newtonian and
relativistic cosmological models in which this is not true. The implication
is that Machian solutions are a subset of all solutions of the EFE,
characterized by some suitable kind of boundary conditions, for examples
Raine's isotropy conditions \cite{raine} resulting from an analysis of
solutions of the remarkable Sciama-Waylen-Gilman (SWG) integral formulation
of the EFE \cite{swg}. This then comes close to supporting Penrose's
proposal \cite{penrose,penrose1} of the necessity of isotropic singularities
at the Big Bang in order that ordinary thermodynamic properties can hold ,
solutions allowing such singularities being studied in some detail by Tod
and Newman \cite{tod}.

On the other hand, some people - notably Barbour \cite{Barbour-Pfister} -
maintain that Einstein's equations already incorporate the full meaning of
what Mach intended, and there is no need for further conditions on the
solutions in order to have a Machian character. This view is imbedded in a
much larger philosophical position supporting the relativity of all
measurements, and suggesting that the passage of time is an illusion - a
position that has received little support from other quarters.

A recent account of the debate on Mach's Principle is given in \cite
{Barbour-Pfister}. This idea has of course been of enormous importance in
the evolution of the theory of general relativity. Three related ideas are
worth mentioning as being of some practical interest.

Firstly, there is the idea of `dragging of inertial frames' occurring in
rotating solutions of the EFE, for example the Kerr rotating black hole
solutions (see section 5 of \cite{Barbour-Pfister}). This then ties Machian
ideas in to the physics of accretion disks in rotating black holes, albeit
it in a rather weak way. Secondly, there have been a series of accurate
experiments relating to Mach's principle (see section 6 of \cite
{Barbour-Pfister}), based on links between Machian ideas and testable
features of local physics. And thirdly, there is the important distinction
made in studies of cosmological perturbations \cite{bardeen} between scalar,
vector, and tensor gravitational modes, and their relation to large scale
structure formation and associated velocities (scalar modes), vorticity
(vector modes), and gravitational waves (tensor modes). Machian issues arise
in each case:\ the degree to which each is affected by very distant as
opposed to local matter. And here an important point arises: tensor modes
are determined by distant matter, because the corresponding characteristic
velocity is the speed of light (gravitational waves propagate at speed $c$),
but in the case of pressure-free matter (the recent universe) the
characteristic velocities of the scalar and vector modes are zero \cite
{charact,jump} - so they are only directly affected by nearby matter. However as
mentioned in the previous section, this result holds only once one has
factored out the background geometry and dynamics - and it is precisely that
full effect (incorporating integrals over all matter) that is the concern of
Mach's principle, not just the perturbation effects around the background
geometry. That is of course both the fatal attraction and flaw in the basic
idea - it is fundamental in its nature (determining the nature of all
inertial effects) but beyond experimental testing in its full extent
(precisely because of that nature). Overall, the idea remains a source of
both inspiration, and irritation , because it is so hard to tie it down
satisfactorily - see page 530 of \cite{Barbour-Pfister} for 21 different
formulations of the idea!

\section{Arrows of Time}

One of the most intractable and fundamental problems in physics is the
relation between reversible fundamental (micro) physics laws, and
irreversible macro-physics effects and associated phenomenological laws,
with the various arrows of time (radiation, thermodynamical, quantum
mechanical, biological for example) dominant in real physical applications 
\cite{whefey}. It seems to be agreed by most that in one way or another this
major discrepancy must reside in the difference between initial conditions
and final conditions in the universe, which effectively disallow half of the
solutions that are in principle allowed by the time-reversible microphysical
equations (see e.g. \cite{ellsci,penrose}). There is however an
alternative view that maybe we are mistaken about the fundamental laws of
physics - maybe they should be formulated in a fundamentally
time-irreversible way \cite{prigogine}. This view has not attracted as much
a following, despite the fact that the irreversibility of physics is clearly
already evident in the quantum measurement process \cite{penrose}, and so is
not solely a macro-physical phenomenon.

It is clear that the arrow of time is closely related to the definition and
evolution of entropy. The relation of a phenomenological (macro) definition
to microscopic properties is obtained via a process of coarse graining \cite
{penr1}, through which the macro-description (given only in terms of
macroscopic variables) explicitly loses information that is available in the
detailed micro description (given in terms of microscopic variables); and
then the basic quantitative question is how many different micro-states
correspond to the same macro-state. A macro-state is more probable if it
corresponds to a greater number of different micro states, and time
evolution will tend to go from a less probable to a more probable state,
defined in this way \cite{penrose}. So far so good; this can be made precise
in terms of Boltzmann's H-theorem, proving the second law of thermodynamics
for suitably defined entropy (defined as an integral over microstate
occupancies) on the basis of microscopic physical laws (see \cite{ehlers}
for a beautiful derivation of this result in the case of relativistic
kinetic theory). But the problem is that this argument applies equally in
both directions of time: it completely fails to determine which is the
forward direction of time. It predicts the entropy will increase in both
directions of time! The only plausible basis so far for making a choice, is
that the local direction of time is determined by boundary conditions on the
physical equations at the beginning (and perhaps also at the end) of the
evolution of the universe \cite{ellsci,penrose}. This seems to be the
logical explanation - but how this master arrow of time uniquely determines
the directions of each of the separate physical and biological arrows still
needs convincing explication. This is one of the most important unsolved
problems in present day physics, for it represents a major gulf between the
macro-physics that dominates everyday life, and the microphysics that (on a
reductionist viewpoint) is supposed to explain that macrophysics. The small
amount of attention paid to it is presumably due to the intractability of
the problem.

The important application of this idea is in terms of the questions raised
by Roger Penrose regarding the idea of an inflationary universe and the
initial conditions for physical fields required in order to obtain a
macroscopic second law of thermodynamics that agrees with the direction of
the inflationary expansion \cite{penrose,penrose2}. He argues that this will
work only if the initial state is very special relative to a random state,
where many black holes might occur with huge initial entropies. Hence the
universe must start off from a highly special state corresponding to small
Weyl tensor magnitudes and an initially isotropic expansion, if the arrow of
time is to work out in terms of the gravitational field - in contrast to the
random initial conditions suggested by many theories of inflation, for
example chaotic inflation \cite{linde}. It is something of a mystery that
this argument does not seem to be given the attention it deserves. This may
have more to do with the sociology of science than the merits of the theory.
The strength of his remarks is strengthened by an appreciation of the
trans-Planckian problem of some inflationary cosmology models \cite
{transplanck}: namely that the perturbations that are supposed to lead, on
the inflationary view, to some large scale astronomical structures, can lie
deep within the Planck era, where the linearised gravitational theory that
is normally used simply does not apply; and indeed we might expect some kind
of space-time foam description to be accurate. Amplifying such inhomogeneity
through inflationary expansion to large scales would lead to anything but a
smooth structure. This issue remains unresolved.

\subsection{Gravitation and Entropy}

A crucial element that is missing from our understanding of the growth of
structure in the expanding universe, is a suitable definition of the entropy
of an arbitrary gravitational field. Of course a huge amount has been
written on the concept of entropy of a black hole \cite{bh}; but that does
not deal with the fundamentally important question of what entropy is
associated with inhomogeneities of the gravitational field when no black
holes are involved.

The importance of this topic is that it underlies the growth of astronomical
structure (galaxies, stars, and planets) in the universe, which in turn
allows our own existence. If you take the statements about entropy in almost
every elementary textbook, and indeed most advanced ones, they are
contradicted when the gravitational field is turned on and is significant 
\cite{ellamj}. For example, in the famous case of the gas container split
into two halves by a barrier, with all the gas initially on one side, the
standard statement is that the gas then spreads out to uniformly fill the
whole container when the barrier is removed, with the entropy
correspondingly increasing. But when gravitation is turned on, the final
state is with all the matter clumped in a blob somewhere in the container,
rather than being uniformly spread out. This is related to the issue of the
negative specific-heat behaviour of gravitational systems, and the
associated `gravithermal catastrophe' \cite{gravithermal} .The question then
is whether there is a definition of entropy for the gravitational field
itself (as distinct from the matter filling space-time), and if so if the
second law of thermodynamics applies to the system when this gravitational
entropy is taken into account. The answers are far from obvious. Dyson \cite
{dyson} claims there is no such entropy, while Penrose \cite
{penrose1,penrose} claims there is, and that it is related to some integral
of the Weyl tensor on spacelike surfaces. A flurry of recent work has given
local definitions of gravitational entropy in relation to the idea of
`holography' in the expanding universe \cite{holography}, but that
literature has not shown how this entropy behaves as structure is formed.
Tavakol and Ellis (unpublished)\ have conjectured that gravitational entropy
is related to the spatial divergence of the electric part $%
E_{ac}=C_{abcd}u^bu^d$ of the Weyl tensor measured by an observer moving
with 4-veocity $u^a$, which certainly seems to have some of the desired
properties, because of the Bianchi identity

\[
\nabla ^aE_{ab}=\frac 13\nabla _b\rho 
\]
relating that divergence to the spatial gradient of the matter density $\rho 
$ (in linearised gravitational theory), where $\nabla $ is the covariant
derivative operator orthogonal to $u$. However that work is incomplete - we
have not succeeded in showing some suitable function of this quantity has
all the desirable properties for a gravitational entropy. Of course the
problem is related to the well-known difficulties in obtaining local
definitions of the mass of an isolated system in general relativity. Recent
work by Ashtekar \cite{ashtekar} may open the way here, and that in turn
might provide a new approach to the entropy issue, because of the well-known
relations between entropy and energy on the one hand, and between mass and
energy on the other.

This issue remains one of the most significant unsolved problems in
classical gravitational theory, for as explained above, even though this is
not usually made explicit, it underlies the spontaneous formation of
structure in the universe - the ability of the universe to act as a
`self-organizing' system where ever more complex structures evolve by
natural processes, starting off with structure formed by the action of the
gravitational field. If solved in a generic way, it would also presumably
lead to another view of the nature of black hole entropy, for it would
certainly have to tie in to that concept in a congruent way. It would
presumably also relate to the idea of the coarse graining of the
gravitational field, and hence to the whole difficult averaging problem in
general relativity theory \cite{GR10}, which may not be approachable in a
covariant and gauge invariant manner (cf. \cite{jelle,matell}).

Given a suitable definition of gravitational entropy and a proof that it has
the required local properties, the further issue that will remain is how
this entropy can tie in to the master arrow of time given by the expansion
of the universe. At one level, implicit answers to this are given by the
standard theory of structure formation in the expanding universe \cite
{inflate1,inflate2}; however, firstly this has not been explicitly related
to a general theory of gravitational entropy, and secondly, as reported
above, Penrose has raised major questions about whether the usual
explanation in terms of an inflationary model will in fact work, when one
looks at its implications for entropy \cite{penrose2}.

\section{Finite Infinity and Local Physics}

In looking at the evolution of `isolated systems', such as the solar system,
our Galaxy, or the local group of galaxies, it is common to use the idea of
asymptotic flatness as the setting for the local system, and to put boundary
conditions on local physical fields `at infinity'. This has been taken to a
very high level of sophistication \cite{ehlers book}, particularly by use of
Penrose's concept of conformal infinity \cite{conformal,hawell}. But
that description makes it very difficult to look at the relation between
such `isolated systems' and the universe in which they are imbedded,
precisely because it ignores the structure of that universe. In the real
universe, there will probably be no asymptotically flat region at or `near'
infinity, for the real universe is almost certainly not asymptotically
Minkowskian at very large distances, and indeed it may not be spatially
infinite.

This led me some years ago to ask the question: `How far away is an
effective `infinity' to use in discussing boundary conditions for local
physical systems of this kind?' Since that was written, answers have been
given for the solar system \cite{solar syst} ( between 1.5 and 3 light
years) and for the local group of galaxies \cite{vdbergh} (about 1.2 Mpc).
So the obvious proposal \cite{GR10} is that we should put boundary
conditions on all fields at that distance, rather than at infinity itself,
leading to the concept of a `finite infinity' $\mathcal{F}$: a smooth
timelike surface at a large but finite distance $r_{*}$ from the centre of
the system considered, separating it from the surrounding universe, and
lying in an almost-flat space-time region at that distance. Then incoming
and outgoing radiation conditions can be imposed on that surface $\mathcal{F}
$, rather than at infinity or conformal infinity $\mathcal{I}$ as is usual 
\cite{ehlers book}. One can then examine the physics of the interaction
between the interior and exterior regions by relating each to energy,
momentum, matter, and any fields that cross this surface. This timelike
surface will, in conformally flat coordinates, be an extremely long, thin
tube (the radius of the tube for the solar system will be about one light
year, but we will want to consider incoming and outgoing radiation for many
hundreds of years when we look for example at the stability of the solar
system).

This idea has not been fully developed yet, except to some degree in the
case of the `swiss-cheese' models of spherical inhomogeneities of the
gravitational field \cite{swisscheese}, but raises many interesting issues.
One needs a clear definition of when the bounding surface lies in an
`asymptotically flat' region around the isolated system but at a finite
distance from it, which can be defined by requiring (i) existence of
suitable local almost-flat coordinates in a neighbourhood of the surface,
and (ii) limits on the amount of matter and radiation present there. In the
cosmological context, (iii) the velocity of the surface must not be greatly
different from that of the cosmic background radiation, as otherwise that
radiation will be experienced as high intensity radiation as it crosses $%
\mathcal{F}$, and so the system will no longer be `isolated'. As regards
(i), we might for example require the existence of an atlas including a set
of local coordinates $x^i=(t,r,\theta ,\phi )$ in a region $U$ $=I_1\times
I_2$ $\times S_2$ where $I_1=(t_1,t_2)$, $I_2=(r_1,r_2)$ are finite
intervals with $r_{*}\in I_2$, such that (a) in $U$, the metric takes the
almost-flat form

\begin{equation}
ds^2=-dt^2+dr^2+r^2(d\theta ^2+\sin ^2\theta \,d\phi
^2)+h_{ab}(x^j)dx^adx^b,\;|h_{ab}|<<|g_{ab}|  \label{asflat}
\end{equation}
(b) $U$ surrounds the finite system (either the coordinates $x^i$ in fact
cover a larger region of space-time, including the isolated system and it
centre, but is only asymptotically flat in the region $U$; or other
coordinates cover the interior of $U$ and include the system), and (c) the
exterior universe lies outside $U$ (usually this will require a separate
coordinate system).

Such a surface will not be uniquely defined; if one such surface exists
there will be an allowed family of such surfaces, related to each other by a
rather large group of transformations - from (i), they must not be too far
out, nor too far in; from (iii), their velocities must not be too different
from each other; and they should be suitably smooth; apart from this, they
can be chosen arbitrarily. One might allow a density discontinuity across
this surface, as in the case of the Swiss-cheese models, in order to match a
fluid-filled exterior to a vacuum interior; but that will need to be handled
with caution (and would require a modification of the formulation of
condition (i) given above). In that case, the exterior boundary must be
chosen as comoving. The obvious choice of velocity for this family of
surfaces will be that velocity field which sets the microwave background
anisotropy dipole to zero; this will be comoving with the matter to a high
degree of accuracy, and the degree to which this is not true will be an
important statement about the nature of the interaction of the chosen system
with the universe.

To develop the dynamical implications of this idea, one needs to develop the
boundary value problem for all the fields involved, generalizing the usual
proofs to a surface with both spacelike and timelike segments. In a major
step forward, Helmut Friedrich and G Nagy \cite{Friedrich} have developed
this boundary value problem for the case of the vacuum gravitational field.
They did not give a physical interpretation to the mathematical variables
defined in their theorems, but such an interpretation should be forthcoming
in terms of a preferred family of observers associated with the allowed
family of surfaces $\mathcal{F}$, particularly when the outer region is
defined in a cosmological context. They also did not consider the
implications when this surface lies in an almost flat region, as for example
specified by Eqn.(\ref{asflat}) above. The further challenge then will be to
use a suitable definition of mass defined from data on this surface (cf.
e.g. \cite{mass}) and then generalize to this setting the results by Bondi
et al, Sachs, and Newman and Penrose relating incoming and outgoing
radiation at infinity to mass loss and the `news function' \cite{penrin}.
Furthermore the famous positive mass theorems \cite{posmass} should also be
generalized to this case. These will be more complex than the existing
proofs with the limit taken at infinity, but in my view the added
complication will not be gratuitous, but will be essential to the physics of
the situation. Additionally, the impact of the constraints on the allowed
incoming and outgoing radiation needs careful analysis \cite{jump}. This
will then provide a rigorous setting for discussing the real physics of the
interaction of the interior and the exterior regions, with physical
estimates of the magnitude of incoming and outgoing radiation that crosses
this surface, and limits on these quantities in order that the system can
indeed be regarded as physically isolated.

This will then be the proper setting in which to look at the relation of
local physics to cosmology, and hence to look at the other questions
discussed in this paper - the Mach and Arrow of Time issues, for example
(this kind of description is already implicit in the Olber's type of
calculations that are carried out). A useful step in this regard is a paper
by Hogan and Ellis \cite{hogell} determining the radiation part of the
electromagnetic field at such a surface at a finite distance from the
source, in a flat background spacetime (and hence, by conformal
transformation, in FLRW geometries). This is in contrast to usual
definitions of the radiation part of the electromagnetic field, which use a
limit at infinity. It would be useful to carry out similar calculations for
the gravitational field, probably best based on the SWG\ integral formalism
mentioned above \cite{swg}.

This may also be the best setting for numerical calculations for `isolated
systems', which often talk about `integrating to infinity', but in most
cases do nothing of the sort\footnote{%
But see H\"{u}bner \cite{hub} for a conformal method that integrates to null
infinity.}. As in the rest of theoretical physics, it would be advantageous
to have a theoretical framework that corresponds more closely to actual
calculations - namely an integration to a surface at a finite distance from
the centre of coordinates. It is usual to make that surface a null surface;
the suggestion here is that it would be better to make it timelike,
corresponding to the region in the real universe where the exterior is
physically separated from the local system. A null surface does not work
well in this context, hence the importance of Friderich's and Nagy's work on
the mixed initial value problem. It might be useful seeing if a conformal
version of their theory would be suitable for numerical work.

Overall, the point is that no system can be completely isolated; the context
suggested here allows one to monitor the degree to which any local system is
indeed isolated, and to examine the nature of its interaction with the
external world - that is, with the rest of the universe. The criterion for
`isolation' will be a real physical one in terms of limits on incoming and
outgoing effects (matter, radiation of all kinds, and tidal forces) across
the separating surface $\mathcal{F}$, rather than statements on limits at an
unattainable infinity, as has been customary up to now (for example in
studies of gravitational radiation and of the Hawking effect). In my view
this will make the analysis much more useful, and genuinely physical based
in terms of relating to real estimates of the magnitude of these effects.
The difference corresponds to the transition in mathematics between calculus
based on infinite limits, and analysis based on $\varepsilon $ and $\delta $
bounds. I believe it will have the same kind of beneficial effects.

\subsection{Possibility of Newtonian Physics}

A particular interesting point then, is what kinds of conditions on a
surface $\mathcal{F}$ surrounding a local physical system will be required
in order that that system can validly be described in terms of Newtonian
physics (cf. \cite{vanell} for an examination of the associated consistency
conditions from another point of view). The point here is that too much
interference from the outside will prevent a good Newtonian limit existing,
for example a local system imbedded in a universe where high-intensity
gravitational waves abound will not have a good Newtonian description. Thus
there will be limits on the particles and gravitational waves crossing any
bounding surface like $\mathcal{F},$ in order that such a description be
possible.

It seems likely that existence of a surface $\mathcal{F}$ in an almost flat
region at a finite distance, such as suggested above, may be sufficient to
show the possibility of Newtonian-like behaviour (it will not guarantee it,
since black holes may form inside such a surface). It is possible that such
existence can \emph{only} occur if the exterior universe is reasonably
similar to a standard Friedmann-Lema\^{\i}tre-Robertson-Walker (`FLRW')
model, and with almost co-moving boundaries chosen in the exterior region.
That is, it seems likely that if the universe is not suitably close to a
FLRW\ model, then no suitable surface $\mathcal{F}$ will exist and a
Newtonian limit will not be attainable locally in such a universe. This is
an idea that needs to be checked. Whether this is the case or not, the
condition here is a significant one that needs investigation: when does a
universe allow Newtonian-like behaviour in local regions? This is clearly an
important aspect of how global structure affects the nature of local
physics. This corresponds to the dual micro-question: when does a quantum
system behave in an almost classical way, and what kind of cosmologies allow
the emergence of classical regimes from the early quantum domain?

\section{Initial Conditions and Local Physics}

The setting just described enables one to distinguish matter and radiation
that crosses the surface $\mathcal{F}$, in either an ingoing or outgoing
direction, from matter which does not do so, and hence is associated with
the isolated system on a long-term basis. Such matter relates the `isolated
system' to local initial conditions at the same position (as defined by
comoving with matter) but at very early times, rather than to exterior
fields at later times. This also applies to scalar and vector gravitational
modes, for the crucial point was made above: the characteristics for these
modes are timelike curves rather than null surfaces, and how they behave is
determined by conditions near our world line at very early times. Thus the
true domain of dependence of such modes is not the usual domain of
dependence bounded by null curves \cite{hawell}, but rather a much smaller
region close to the matter world lines.

The practical issue is that many local conditions at the present time are
determined by conditions near our world line in the very early universe \cite
{ell71}, and this applies particularly to element abundances (based on
nucleosynthesis) and baryon abundances (based on baryosynthesis). Each are
relics of non-equilibrium phases in the early universe's history. A specific
set of such relics results from specific initial conditions, and determines
the nature of what can exist locally at the present time (stars, planets,
and living beings have to be constructed out of whatever matter is present
locally). The power of the assumption of equilibrium physics in the early
universe is that the nature of these relics is very largely independent of
what existed at very early times, for whatever one might feed in will get
transformed into an equilibrium mixture (this is the cosmological version of
the famous statement that the nature of black holes is independent of what
is put into them). Only conserved quantities will survive unchanged; and
there are very few of those in the extreme conditions of the very early
universe.

\section{The Existence of Life}

A key issue for the future is to clarify in more detail the relation of the
nature of the universe to the existence of life, both in terms of initial
conditions, and of the nature of the laws of physics. This is the highly
contested terrain of the \textit{Anthropic Principle}. Strangely, Dennis
wrote rather little on this, but it has been a very active area and
certainly is a legitimate concern within the broad terrain under discussion.

This concept has been regarded with considerable suspicion by many because
of some rather unguarded or ill-thought out statements regarding the nature
and application of the anthropic principle. There is a clear distinction
between the \textit{Weak Anthropic Principle} (`WAP') from the \textit{%
Strong Anthropic Principle} (`SAP') \cite{bartip,carter}. The former is an
unobjectionable selection principle (`we can only view the universe from
space-time regions that allow our existence'), while the latter is a highly
disputable philosophical claim (`the universe must allow the existence of
life'), argued on a number of different grounds, for example the need for
observers to exist in order that quantum theory can make sense. The problem
then is that firstly, some papers seem to argue the SAP case by an inversion
of normal logic, for example the statement by Collins and Hawking at the end
of an important study of Bianchi cosmologies that `the universe must exist
because we are here' \cite{colhaw}; and secondly there have been some
attempts to stretch the concept into quite undefensible territory \cite
{ellanth}, in particular Barrow and Tipler's \textit{Final Anthropic
Principle} (`FAP'): `life not only must exist, but once it has come into
existence must continue to exist until the end of the universe' \cite{bartip}%
. This dubious proposition led Gardner in a famous review to refer to the 
\textit{Completely Ridiculous Anthropic Principle} (`CRAP').

These extremes are unfortunate, because they have obscured important
arguments regarding the nature of the laws of physics and boundary
conditions in the universe necessary to the existence of life. If one leaves
aside the contentious claims, such as those mentioned above, one is left
with an important selection principle that may indeed be essential both in
terms of trying to explain the value of the cosmological constant \cite
{weinberg} and in relating concepts such as chaotic inflation to
observations \cite{linde}. There is also an intrinsic interest in charting
out what variations in physical laws will allow life to survive \cite
{tegmark}. Thus in my view this is indeed an important part of the range of
issues discussed in this article, and the need is to strictly separate out
the unexceptional range of concerns that can be classed as WAP\ issues, and
their use for example in terms of relating observations to chaotic
inflation, from the whole range of controversial concerns raised under the
SAP banner. These relate to fundamental metaphysical issues that are of a
different nature than the issues pursued here; they are of course important,
but are of a different character and need a different kind of discussion of
a philosophical and metaphysical character \cite{ellphil,ellunique} rather than
relating to the strictly physical issues discussed in this paper.

The WAP issues are very much part of the theme of this paper. As they relate
the rest of the arguments to questions concerning human life and existence,
they are of major importance and interest to us as human beings as well as
scientists.

\section{Conclusion}

As well as the major way that microphysics affects macrophysics in a
`bottom-up' way, there are many themes whereby there is a `top-down' action
of the cosmos as a whole on local physical systems. Some unsolved problems
of physics may be related to this theme, in particular the `arrow of time'
issue which is still a major puzzle for theoretical physics. Many of the
themes discussed here have practical applications in terms of being related
to tests of cosmological theories, precisely because if the universe has an
influence on local systems, then observing local systems tells us something
about the universe. This range of themes remains of interest today; there is
still interesting work to be done on them. These issues were amongst the
driving forces of Dennis' career, and he presented them with force on many
occasions.

I thank Henk van Elst and Roy Maartens for helpful comments on earlier
versions of this paper.

\end{document}